**Fundamentals of legislation for autonomous artificial intelligence systems**

Romanova Anna, Ph.D.

Moscow Institute of Physics and Technology (National Research University), postgraduate student

Moscow, Russian Federation

romanova.as@phystech.edu

ORCID 0009-0004-4649-6037

Annotation

The article proposes a method for forming a dedicated operational context in course of development and implementation of autonomous corporate management systems based on example of autonomous systems for a board of directors. The significant part of the operational context for autonomous company management systems is the regulatory and legal environment within which corporations operate. In order to create a special operational context for autonomous artificial intelligence systems, the wording of local regulatory documents can be simultaneously presented in two versions: for use by people and for use by autonomous systems. In this case, the artificial intelligence system will get a well-defined operational context that allows such a system to perform functions within the required standards. Local regulations that provide for the specifics of the joint work of individuals and autonomous artificial intelligence systems can create the basis of the relevant legislation governing the development and implementation of autonomous systems.

___

**Начала законодательства для автономных систем искусственного интеллекта**

Романова Анна Сергеевна, кандидат экономических наук.

Московский физико-технический институт (национальный исследовательский университет), аспирант

г. Москва, Российская Федерация

romanova.as@phystech.edu

ORCID 0009-0004-4649-6037

Аннотация

В статье предлагается метод формирования выделенного операционного контекста при разработке и внедрении автономных систем управления корпорациями на примере автономных систем для совета директоров. Значительную часть операционного контекста для автономных систем управления компаниями составляет регуляторная и правовая среда, в рамках которой корпорации осуществляют свою деятельность. С целью создания специального операционного контекста для автономных систем искусственного интеллекта формулировки локальных нормативных документов могут быть одновременно представлены в двух вариантах: для использования людьми и для использования автономными системами. В таком случае система искусственного интеллекта получает четко очерченный операционный контекст, который позволяет такой системе выполнять функции в рамках необходимых эксплуатационных качеств. Локальные нормативные акты, которые предусматривают специфику совместной работы физических лиц и автономных систем искусственного интеллекта, могут стать основой для формирования основ соответствующего законодательства, регулирующего разработку и внедрение автономных систем.

Ключевые слова: автономные системы, искусственный интеллект, операционный контекст, совет директоров, корпоративное управление

**Введение**

Системы искусственного интеллекта (далее ИИ), разрабатываемые для участия в управлении корпорациями, должны эффективно работать не только с объектами материального мира, но и в правовом поле. "По крайней мере начиная с Лейбница, мечта об исключении человека из спирали юридических рассуждений захватила воображение философов, юристов и (в последнее время) ученых-компьютерщиков" [1]. Идея Лейбница представлена "в его Dissertatio de Arte Combinatoria как Универсальная Математика, теоретическая, формальная система предложений и правил, которая позволила бы разрешить все споры с математической точностью" [1]. Исторически законы создавались и исполнялись людьми. С развитием технологий искусственного интеллекта законы будут исполняться машинами. Уровень точности формулировок, приемлемый для современного человека, намного ниже, чем уровень точности, необходимый для искусственной системы. Существует фундаментальная разница "между человеческими решениями, как социальными конструкциями, и алгоритмическими решениями, как техническими конструкциями" [2]. Математик Стивен Вольфрам в своем выступлении на конференции SXSW 2013 сказал: "вычисления станут центральным элементом почти в каждой области" [3]. Вольфрам считает, что: "теперь мы почти готовы… к вычислительному закону. Где, например, контракты становятся вычислительными. Они явным образом становятся алгоритмами, которые решают, что возможно, а что нет" [3]. Вольфрам предполагает, что для систем искусственного интеллекта необходимо будет принять отдельную конституцию [4]. Однако вопрос: "Что должно быть в такой конституции?" [4] на данный момент остается открытым.

Современные автономные системы ИИ для управления корпорациями уже позиционируются как активное действующее лицо, способное принимать решения на уровне совета директоров, оценивать стратегические опции, предоставлять рекомендации акционерам [5]. Тем не менее, на данный момент, ни в одной стране мира не принято, и даже не разработано законодательство, регулирующее создание и применение таких систем.

Законодатели, разработчики, корпорации, а также их акционеры заинтересованы в том, чтобы автономные системы ИИ могли эффективно применяться в корпоративном управлении, не создавая при этом необоснованных и неуправляемых рисков. Формирование соответствующего законодательства, учитывающего не только теоретические правовые концепции, но технические особенности современных автономных систем ИИ, является жизненно важным и необходимым условием, которое позволит разрабатывать и внедрять законные, этичные, и безопасные автономные системы ИИ в управлении корпорациями.

**Рабочее пространство разработки и операционный контекст**

Наиболее эффективные подходы к разработке и применению гражданских и коммерческих автономных систем в настоящее время выработаны в области создания и использования автономных автомобилей. При создании автономных транспортных средств используется понятие "домен операционного проектирования (ODD)" [6], который "является абстракцией операционного контекста, а его определение является интегральной частью процесса разработки системы" [6]. Международный стандарт J3016 "Таксономия и определения терминов, связанных с системами автоматизации вождения для дорожных транспортных средств" Общества автомобильных инженеров (SAE) определяет домен операционного проектирования как "комбинированные условия эксплуатации, при которых данная система автоматизации вождения (или ее функция) специально разработана для функционирования" [7]. "Необходимо знать операционный контекст, чтобы предоставлять гарантии по эксплуатационным качествам и безопасности" [6]. Соответственно, "необходимый уровень безопасности гарантируется только в четко определенном и испытанном домене операционного проектирования" [6]. Британский институт стандартов (BSI) указывает, что "ключевым аспектом безопасного использования автоматизированного транспортного средства является определение его возможностей и ограничений и ясная коммуникация об этом конечному пользователю, что приводит к состоянию «информированной безопасности»" [8]. Британский институт стандартов считает, что "первым шагом при установлении возможностей автоматизированных транспортных средств является определение области операционного проектирования (ODD)" [8].

Для автономных систем корпоративного управления частью операционного контекста являются законы и другие нормативно-правовые документы, а также толкование и применение законов и документов другими системами ИИ и людьми. Стивен Вольфрам не единственный, кто пришел к выводу о том, что для эффективной работы автономных систем ИИ необходимо внести существенные изменения в законодательную систему. В отчете Европейской Комиссии по "Этике подключенных и автоматизированных транспортных средств" говорится, что автономные автомобили не смогут буквально соблюдать правила, созданные для людей. Для успешного внедрения беспилотных автомобилей необходимо рассматривать несколько опций: "(a) должны быть изменены правила дорожного движения; (b) автономным автомобилям должно быть разрешено не соблюдать правила дорожного движения; или (c) автономные автомобили должны передавать управление, чтобы человек мог принять решение не соблюдать правила дорожного движения" [9].

Современная юридическая практика уже знает примеры, когда для систем со значительной разницей в мировосприятии создается дополнительный или специальный операционный контекст — это двуязычные контракты. Например, "китайское законодательство требует, чтобы договор о совместном предприятии одобрялся китайскими государственными органами" [10]. Поэтому закономерно, что "договор о совместном предприятии должен быть написан на китайском языке" [10]. В тех же случаях, когда нужен двойной операционный контекст, "договор о совместном предприятии является двуязычным договором: есть один договор, но с двумя разными текстами, один на английском и один на китайском" [10].

Рассматривая ситуации, когда автономная система ИИ должна будет принимать неочевидные для человека решения, многие исследователи по-прежнему формулируют возможные критерии для принятия решения исходя из привычных категорий восприятия человека другим человеком: "раса, религия, пол, инвалидность, возраст, национальность, сексуальная ориентация, гендерная идентичность или гендерное самовыражение" [11]. В знаменитом эксперименте Машина Морали (Moral Machine) исследователи из Массачусетского технологического института (MIT) также использовали только очевидные для человека факторы: пол, возраст, и т.д. [12]. Для систем ИИ, созданных на основе математических алгоритмов и получающих цифровую информацию с помощью разнообразных датчиков, такие критерии являются лишь небольшой частью данных, на основании которых система ИИ рассчитывает решение.

В современном мире уже существуют системы ИИ, которые в автономном режиме решают специфические задачи, напрямую не принимая управленческие решения. Это системы, которые принимают очень быстрые финансовые решения — системы алгоритмического и высокочастотного трейдинга [14]. Для таких систем разработаны специальные регуляторные техники: "раскрытие информации, внутреннее тестирование и системы мониторинга" [14]. Также для таких систем предусмотрены "структурные особенности торгового процесса" [14], т.е. для таких систем создан специальный операционный контекст.

**Формирование операционного контекста для автономных систем управления**

По аналогии со стандартом J3016 "Таксономия и определения терминов, связанных с системами автоматизации вождения для дорожных транспортных средств" для автономных систем управления корпорациями также можно сформулировать понятие домена операционного проектирования: - это комбинированные условия эксплуатации, при которых данная система автоматизации управления (или ее функция) специально разработана для функционирования. Значительную часть операционного контекста для автономных транспортных средств составляют объекты материального мира, однако для автономных систем управления компаниями "регуляторная и правовая среда, в рамках которой корпорации осуществляют свою деятельность, имеет ключевое значение для общих экономических результатов" [14]. Существенную часть операционного контекста для автономных систем управления корпорациями составляют разнообразные нормативные акты. Принципы корпоративного управления G20/ОЭСР устанавливают, что "цели корпоративного управления также формулируются в добровольных кодексах и стандартах, которые не имеют статуса закона или нормативных актов" [14]. С целью создания специального операционного контекста для автономных систем ИИ формулировки локальных нормативных документов могут быть одновременно представлены в двух вариантах: для использования людьми и для использования автономными системами. В таком случае система ИИ получает четко очерченный операционный контекст, который позволяет такой системе выполнять функции в рамках необходимых эксплуатационных качеств.

Основные принципы корпоративного управления, а также базовые функции совета директоров, изложены в "Принципах корпоративного управления G20/ОЭСР" [14]. С целью разъяснения и внедрения Принципов G20/ОЭСР многие страны и компании разрабатывают и применяют собственные, более детальные, кодексы корпоративного управления. Принципы G20/ОЭСР и кодексы корпоративного управления формируют основу, которая является базой для создания операционного контекста для автономных систем управления корпорациями. Рассмотрим несколько примеров формулирования политик для автономных систем ИИ в составе смешанных советов директоров (советов, состоящих из физических лиц и автономных систем ИИ).

*Принцип справедливого отношения ко всем акционерам*

Ключевым принципом корпоративного управления является "справедливое отношение ко всем акционерам" [14]. Понятие справедливого отношения для автономных систем в современной практике формализуется с помощью принципов информированного согласия [9], недискриминации [11], и справедливого статистического распределения рисков [9].

*Информированное согласие*

Современные исследователи систем ИИ приходят к выводу, что "инженеры не имеют морального права принимать этические решения от имени пользователей в трудных случаях, когда ставки высоки" [15]. В отчете по "Этике подключенных и автоматизированных транспортных средств" указывается, что для использования автономных систем необходимо разрабатывать "более тонкие и альтернативные подходы к пользовательским соглашениям" [9] для получения информированного согласия, а не просто подход «соглашайся или уходи»" [9]. Информированное согласие предполагает информирование пользователя о том, как система ИИ будет себя вести в обычных условиях и в критических ситуациях. Аварийные ситуации на дороге, в которых участвовали автопилоты Тесла (Tesla), показывают, что "неясно, были ли бета-тестеры Тесла полностью проинформированы о риске. Знали ли они, что смерть возможна?" [16]. Корпоративные политики, регламенты, и кодексы, в которых описаны основы и правила работы автономной системы ИИ, позволят акционерам и другим заинтересованным лицам выразить информированное согласие на использование такой системы в корпоративном управлении. Поскольку правила корпоративного управления должны исполняться одновременно и директорами - физическими лицами, и автономными системами, они могут быть составлены в двух редакциях - для физических лиц и для автономных систем ИИ:

- политики, регламенты, и кодексы для физических лиц должны регулировать вопросы корпоративного управления, основываясь на мировосприятии физических лиц;
- политики, регламенты, и кодексы для автономных систем ИИ должны регулировать вопросы корпоративного управления, основываясь на показателях, доступных для систем ИИ.

*Недискриминация*

В отчете по "Этике подключенных и автоматизированных транспортных средств" указывается, что необходимо избегать "дискриминационного предоставления услуг" [9] автономными системами. Систему ИИ можно и нужно тестировать на наличие предвзятости, прямой и косвенной дискриминации [17]. Корпоративная политика, составленная для смешанного совета директоров, должна предусматривать какие тесты на непредвзятость должна пройти или регулярно проходить автономная система, чтобы соблюдать правила о непредвзятости, их регулярность, и перечень признаков прямой и косвенной дискриминации.

*Справедливое статистическое распределение рисков*

Принципы G20/ОЭСР устанавливают, что "совет директоров должен соблюдать высокие этические стандарты" [14]. В отчете по "Этике подключенных и автоматизированных транспортных средств" отмечается, что в критических ситуациях "невозможно регулировать точное поведение" [9] автономных систем. Поэтому группа экспертов ЕС предлагает считать поведение автономных систем этичным если "оно органически возникает из непрерывного статистического распределения риска… в целях повышения безопасности… и равенства между категориями участников" [9]. Для справедливого распределения риска современные исследователи пробуют использовать абстрактные величины: "пропорциональная зависимость между скоростями участников дорожного движения и тяжестью вреда может быть установлена независимо от какой-либо этической оценки" [18]. Такой подход не всегда возможен, особенно в случае распределения ограниченных ресурсов. Для аллокации дефицитных медицинских препаратов на практике выработаны несколько алгоритмов: "одинаковое отношение ко всем людям, предпочтение наихудшим случаям, максимизация общих выгод, а также поощрение и вознаграждение социальной полезности" [19]. Поэтому политики, кодексы, регламенты, составленные для смешанного совета директоров, должны раскрывать для акционеров и третьих лиц, каким образом формируются требования к автономным системам в части этики и справедливости, а именно, критерии и показатели для формирования алгоритмов справедливого распределения рисков.

*Мониторинг результатов управленческой деятельности*

Принципы G20/ОЭСР определяют, что "совет директоров, прежде всего, отвечает за контроль результатов управленческой деятельности" [14]. В этом случае могут возникнуть ситуации, когда автономная система ИИ будет оценивать результаты деятельности менеджера — физического лица. В Кодексе корпоративного управления Португалии устанавливается, что "неисполнительные директора должны осуществлять эффективным и разумным образом функцию общего надзора и оспаривания исполнительного руководства" [20]. Более того, по аналогии с концепциями непрерывной отчетности и непрерывного аудита, автономная система ИИ способна осуществлять "непрерывный мониторинг" результатов управленческой деятельности. Концепция "непрерывного аудита" [21] была предложена еще в 1991 году в AT&T Bell Labs для "аудита больших цифровых баз данных" [21]. Система осуществляла "мониторинг и обеспечение в режиме реального времени большой биллинговой системы, фокусируясь на измеряемых данных и идентифицируя с помощью методов аналитики ошибки в данных, что приводит как к контролю, так и к диагностике процесса" [22]. Авторы концепции указывали, что ее внедрение "потребует существенных изменений в характере доказательств, типах процедур, сроках и распределений усилий в аудите" [21], т.е. изменения существующего операционного контекста или создания дополнительного. Концепция непрерывного аудита неразрывно связана с концепцией "непрерывной отчетности" [22]. При создании политик и процедур для автономных систем ИИ компания должна определить, получает ли она значительные конкурентные преимущества при отслеживании транзакций, событий, и информации в режиме реального времени. Политики, кодексы, и регламенты для автономных систем ИИ должны содержать конкретный перечень активностей, источников, и значений для мониторинга эффективности управленческой деятельности.

*Соблюдение законодательства*

Принципы G20/ОЭСР указывают, что в обязанности совета директоров входит "надзор за системой управления рисками и механизмами, предназначенными для обеспечения того, чтобы корпорация соблюдала применимое законодательство" [13]. Современные системы мониторинга регуляторных рисков способны обеспечивать непрерывный мониторинг рисков во многих направлениях одновременно: "соблюдение нормативных требований по борьбе со взяточничеством и коррупцией, соответствие требованиям по борьбе с отмыванием денег, соблюдение нормативных требований в отношении финансовых услуг, оценка рисков текущих и потенциальных деловых партнеров, агентов и поставщиков, риски слияния и поглощения и инвестиций в развивающиеся и глобальные рынки, отраслевые и страновые риски" [23]. Для достижения такого широкого и детального анализа компании создают системы, которые "консолидируют данные из широкого круга мировых источников данных" [23]. Перечень источников данных, которые может собирать и анализировать автономная система может быть очень разнообразным: "ведущие агрегаторы данных, скрининг… в СМИ и/или судебные обзоры, информация о корпоративной структуре и операционной деятельности, владения третьих лиц и акционеры" [23]. Корпоративные политики, кодексы, и регламенты, составленная для автономных систем ИИ в составе смешанного совета директоров должны содержать конкретный перечень источников и расписание обновлений информации.

*Соблюдение интересов третьих лиц*

При раскрытии обязанностей советов директоров Принципы G20/ОЭСР определяют, что "ожидается, что они будут учитывать и справедливо относиться к интересам заинтересованных сторон, в том числе сотрудников, кредиторов, клиентов, поставщиков и затронутых сообществ" [14]. В отчете Европейской Комиссии по "Этике подключенных и автоматизированных транспортных средств" предлагается, что автономные системы должны "адаптировать свое поведение к менее защищенным участникам дорожного движения, вместо того чтобы ожидать, что эти пользователи будут сами адаптироваться" [9]. Также предлагается, что автономные системы "должны быть разработаны таким образом, чтобы принимать активные меры для продвижения инклюзивности" [9]. Исследователи из Мюнхенского технического университета предлагают включать в алгоритм справедливого распределения рисков специальные параметры для менее защищенных пользователей [18]. Корпоративные политики, кодексы, и регламенты, составленные для автономных систем ИИ в составе смешанного совета директоров, должны содержать параметры и соответствующие веса, которые необходимо учитывать при рассмотрении интересов третьих лиц.

*Информированность, добросовестность, осмотрительность, и заботливость*

Принципы G20/ОЭСР устанавливают, что "члены совета должны действовать в условиях полной информированности, добросовестно, с должной осмотрительностью и заботливостью, в наилучших интересах компании и акционеров." [14]. Для систем ИИ информированность формализуется в перечне и объеме необходимых источников и данных, а также регулярности обновления источников, данных, алгоритмов и моделей. При значительном объеме транзакций для человека невозможно установить обязанность проверять каждую операцию и в любой момент времени. Система ИИ может проверять транзакции в режиме реального времени, либо с определенным интервалом [24], [25]. Также система ИИ может проверять все операции, либо только определенные [24], [25]. Для системы ИИ возможно установить и количество, и перечень источников, которыми она будет пользоваться [24]. По аналогии с концепциями непрерывной отчетности и непрерывного аудита компания может рассмотреть опцию внедрения "непрерывной информированности" и "непрерывного мониторинга". Корпоративные политики, кодексы, и регламенты, составленные для автономных систем ИИ в составе смешанного совета директоров должны отвечать на следующие вопросы: каков перечень источников информации, какова регулярность обновления источников, данных, и моделей, каков каталог необходимых активностей.

*Назначение на должность директора*

Принципы G20/ОЭСР предлагают при назначении физического лица директором учитывать его "соответствующие знания, компетенцию и опыт" [14]. Например, в Кодексе корпоративного управления Саудовской Аравии указано, что информация о номинируемых на должность директора кандидатах должна раскрывать "опыт, квалификацию, навыки и их предыдущие и текущие места работы и членства" [26]. Также предусмотрено требование, что кандидат "должен иметь академическую квалификацию и надлежащие профессиональные и личные навыки, а также соответствующий уровень подготовки и практический опыт" [26]. В настоящее время многие крупные банки запрещают сотрудникам использовать систему ChatGPT в бизнес целях в связи с ее "неточностью и законодательными проблемами" [27]. Корпоративные политики, кодексы, и регламенты, составленная для автономной системы в составе смешанного совета директоров должны отвечать на следующие вопросы: по каким параметрам выбирается система ИИ, какие тесты или экзамены она должна пройти.

*Оценка деятельности СД*

В Принципах G20/ОЭСР указывается, что "советы директоров должны проводить регулярную оценку своей деятельности и определять, обладают ли они необходимым сочетанием опыта и компетенций" [14]. Принципы G20/ОЭСР предполагают, что "с помощью тренингов" [14] члены СД могут поддерживать необходимый уровень знаний. Для автономной системы необходимы не курсы и тренинги, а регулярное обновление данных и алгоритмов. Поэтому корпоративные политики, кодексы, и регламенты, составленные для автономных систем ИИ в составе смешанного совета директоров должны отвечать на следующие вопросы: как часто, в каком объеме, на базе каких источников должно проводиться обновление алгоритмов и данных.

*Сотрудничество*

Для смешанного совета директоров необходимо предусмотреть методы эффективной коммуникации между автономными системами ИИ и другими заинтересованными лицами (директорами, акционерами, менеджерами, сотрудниками, и т. д.). Физические лица обычно работают в соответсвии с распорядком трудового дня. Такая периодичность оправдана для распределения эффективной нагрузки для физических лиц, но не имеет смысловой нагрузки для определения режима работы автономной системы ИИ, которая может работать круглосуточно. Также меняется само понятие деловых встреч и эффективной коммуникации. Автономная система ИИ будет использовать для общения цифровой интерфейс: "чат-боты (например, разговорный ИИ через аудио или текст), визуальные голограммы, виртуальную или дополненную реальность" [28].

Лучшие практики формулирования дополнительного операционного контекста для автономных систем ИИ в дельнейшем могут быть обобщены и использованы в законодательной деятельности. Можно было бы предположить, что физические лица также предпочтут использовать более точные формулировки, созданные для автономных систем, но физические лица не смогут обработать необходимый объем данных.

**Заключение**

В настоящее время законодатели, разработчики, и исследователи в области искусственного интеллекта стоят перед выбором: нужно ли создавать специальные условия для функционирования автономных систем искусственного интеллекта или они могут функционировать в том же операционном контексте, что обычные люди. Иными словами, можно ли рассматривать автономные системы как автомобиль, который может использоваться на автотрассе общего пользования, или это скорее поезд, самолет, или ракета, и для эффективного использования таких систем нужна выделенная инфраструктура.

Анализ аварийных ситуаций на транспорте [29] однозначно показывает, что несмотря на несопоставимую мощность и скорость, авиатранспорт и железнодорожный транспорт в несколько раз безопаснее автомобилей. Огромную и безопасную, по меркам автомобильного транспорта, скорость и мощность самолетам и поездам позволяет достигать выделенная инфраструктура: железнодорожные пути, вокзалы, аэропорты и воздушные коридоры. Такой же принцип повышения безопасной эффективности за счет выделенной инфраструктуры уже частично применяется в области алгоритмического и высокочастотного трейдинга и может применяться для других автономных систем ИИ.

В области корпоративного управления значительная часть инфраструктуры создается в виде внутренних нормативных актов компании. Локальные нормативные акты, которые предусматривают специфику совместной работы физических лиц и автономных систем ИИ, могут стать основой для формирования основ соответствующего законодательства, регулирующего разработку и внедрение автономных систем ИИ.

**Литература**

**Сведения об авторе**

Романова Анна Сергеевна — эксперт в области цифровой трансформации, к.э.н., MBA, LL.M, ALM, FCCA, свыше 15 лет опыта работы в крупнейших российских и международных компаниях. Аспирант МФТИ по направлению "Искусственный интеллект и машинное обучение", Москва, Российская Федерация, e-mail: romanova.as@phystech.edu.

**About the author**

Romanova Anna is an expert in the area of digital transformation, Ph.D., MBA, LL.M, ALM, FCCA, over 15 years of experience in the largest Russian and international companies. Postgraduate student at MIPT in the field of "Artificial Intelligence and Machine Learning", Moscow, Russian Federation, e-mail: romanova.as@phystech.edu.

**Вклад автора**

Романовой А. С. предложена концепция выделенного операционного контекста для автономных систем искусственного интеллекта. Разработанный метод использован для формирования выделенного операционного контекста при разработке и внедрении автономных систем управления корпорациями на примере автономных систем ИИ для советов директоров.

**Конфликт интересов**

Автор заявляет об отсутствии конфликта интересов.